\documentclass[preprint,showpacs,preprintnumbers,amsmath,amssymb]{revtex4}

\usepackage{graphicx}
\usepackage{dcolumn}
\usepackage{bm}

\begin{document}

\preprint{}

\title{Hadronic Decays Involving Heavy Pentaquarks}

\author{Xiao-Gang He}
 \altaffiliation[On leave of absence from]{ Physics Department, National Taiwan University.}
 \email{hexg@phys.ntu.edu.tw}
 \affiliation{Department of Physics, Peking University, Beijing,
 China}
\author{Xue-Qian Li}%
 \email{lixq@nankai.edu.cn}
\affiliation{%
Department of Physics, Nankai University, Tianjin, China
}%


\date{\today}

\begin{abstract}
Recently several experiments have reported evidences for
pentaquark $\Theta^+$. H1 experiment at HERA-B has also reported
evidence for $\Theta_c$. $\Theta^+$ is interpreted as a bound
state of an $\bar s$ with other four light quarks $udud$ which is
a member of the anti-decuplet under flavor $SU(3)_f$. While
$\Theta_c$ is a state by replacing the $\bar s$ in $\Theta^+$ by a
$\bar c$. One can also form $\Theta_b$ by replacing the $\bar s$
by a $\bar b$. The charmed and bottomed heavy pentaquarks form
triplets and anti-sixtets under $SU(3)_f$. We study decay
processes involving at least one heavy pentaquark using $SU(3)_f$
and estimate the decay widths for some decay modes. We find
several relations for heavy pentaquarks decay into another heavy
pentaquark and a $B (B^*)$ or a $D(D^*)$ which can be tested in
the future. $B$ can decay through weak interaction to charmed
heavy pentaquarks. We also study some $B$ decay modes with a heavy
pebtaquark in the final states. Experiments at the current $B$
factories can provide important information about the heavy
pentaquark properties.
\end{abstract}

\maketitle


\section{Introduction}

Recently several experiments have reported evidences for
pentaquarks $\Theta^+$ and other states\cite{1}, although there
are also experiments reported null results\cite{1a}. The
$\Theta^+(1540)$ pentaquark has strangeness $S=+1$ and has quark
content $udud\bar s$. This particle is an isosinglet which is a
member of the anti-decuplet multiplet\cite{2} in flavor $SU(3)_f$
symmetry. At present there is very limited experimental
information on the detailed properties such as the decay width,
spin and parity. Several models have been proposed to accommodate
these states\cite{2,2a,3,4,4a,4aa,4b,5,6,6a,7}.

Replacing the $\bar s$ in $\Theta^+$ by a heavy quark such as a
$\bar c$ or a $\bar b$, it is also possible to form bound heavy
pentaquark states\cite{3,4,4aa,5,6,6a1,6a,8,8a,9,10}, $\Theta_c$
or $\Theta_b$.  When implementing them into $SU(3)_f$, in the
model of Jaffe and Wilczek\cite{3} where pentaquark $\Theta^+$ is
composed of two $(ud)$ diquarks with spin-0 and an $\bar s$ quark,
heavy pentaquarks form a fundamental representation of $SU(3)_f$
triplet $R_{c,b}$ (the sub-indices $c$ and $b$ indicate whether
the pentaquark is formed with a $\bar c$ or a $\bar b$), and an
anti-sixtet $S_{c,b}$\cite{5}. Discovery of these states and study
of their properties can provide important information about the
inner structure of matter. H1 collaboration has recently reported
observation of a narrow resonance in\cite{H1} $D^{*-} p$ and
$D^{*+}\bar p$ in inelastic $eP$ collisions at center-of-mass
enegies of 300 GeV and 320 GeV at HERA. This resonance has a mass
of $3099\pm 3(stat) \pm 5(syst)$ with a Gaussian width of $12\pm
3$ MeV and can be interpreted as evidence for $\Theta_c$. If this
observation is confirmed, there should be other related states
exist. In this paper we study some properties of decay processes
involving at least one heavy pentaquark using $SU(3)_f$ flavor
symmetry.

The triplet $R_{c,b}$ transforms under $SU(3)_f$ in a similar way
as the light quark $(u,d,s)$ triplet. To indicate this fact and
also to distinguish the pentaquark triplet from the quark one, we
use capital $U, D, S$ to indicate the elements in $R_{c,b}$. For
the anti-sixtet $S_{c,b}$, we use $T_{c,b}$, $N_{c,b}$ and
$\Theta_{c,b}$ to indicate the isospin triplet, doublet and
singlet in $S_{c,b}$, respectively. We have

\begin{eqnarray}
&&R_{c} =(R_{c,i})= (U^0_{c},D^-_{c}, S^-_{c}),\nonumber\\
&&S_c = (S^{ij}_c) = \left ( \begin{array}{lll}
T^{--}_{c}&T^-_c/\sqrt{2}&N^-_c/\sqrt{2}\\
T^-_c/\sqrt{2}&T^0_c&N^0_c/\sqrt{2}\\
N^-_c/\sqrt{2}&N^0_c/\sqrt{2}&\Theta_c^0
\end{array}
\right ),\nonumber
\end{eqnarray}
\begin{eqnarray}
&&R_{b} =(R_{b,i})= (U^+_b,D^0_b, S^0_b),\nonumber\\
&&S_b = (S^{ij}_b) = \left ( \begin{array}{lll}
T^{-}_b&T^0_b/\sqrt{2}&N^0_b/\sqrt{2}\\
T^0_b/\sqrt{2}&T^+_b&N^+_b/\sqrt{2}\\
N^0_b/\sqrt{2}&N^+_b/\sqrt{2}&\Theta_b^+
\end{array}
\right ).
\end{eqnarray}

The quark contents of these particles are,

\begin{eqnarray}
&&U^{0,+}_{c,b} = (udus)(\bar c, \bar b),\;\;D^{-,0}_{c,b} =
(udds)(\bar c, \bar b),\;\;S^{-,0}_{c,b} = (dsus)(\bar c,\bar b),\nonumber\\
&&T^{--,-}_{c,b} = (dsds)(\bar c,\bar b),\;\;
T^{-,0}_{c,b} = (dsus)(\bar c,\bar b),\;T^{0,+}_{c,b} = (usus)(\bar c,\bar b),\nonumber\\
&&N^{-,0}_{c,b} = (udds)(\bar c, \bar b),\;\;N^{0,+}_{c,b} =
(udus)(\bar c,\bar b),\;\;\Theta^{0,+}_{c,b} = (udud)(\bar c, \bar
b).
\end{eqnarray}
The $U_{c,b}$, $D_{c,b}$ and $N_{c,b}$ particles have $S= -1$,
$S_{c,b}$ and $T_{c,b}$ particles have $S = -2$, and
$\Theta_{c,b}$ particles have $S = 0$.

In the diquark model of Ref.\cite{3}, $S_{c,b}$ have positive
parity, whereas $R_{c,b}$ have negative parity since there is no
P-wave excitation between the diquarks. In our study we emphasis
on the flavor $SU(3)_f$ properties, the conclusions can be applied
to both parity situations ``+'' or ``-'' for both $R_{c,b}$ and
$S_{c,b}$.

\section{Heavy Pentaquark Strong Decay Couplings}

Whether heavy pentaquarks can have strong decay modes depends on
their masses. With $SU(3)_f$ symmetry, particles in each multiplet
are supposed to have the same mass. Quark model estimates for the
heavy pentaquark have been carried out by several groups. In the
diquqrk model, the $\Theta_{c,b}$ masses are estimated to
be\cite{3} 2710 MeV and 6050 MeV, respectively, which is below the
strong $pD$ and $nB$ decay threshold. A lattice calculation in
Ref.\cite{4aa}, gives $m_{\Theta_c}$ about 3.5 GeV.

Removing the P-wave excitation energy, which is estimated using
the mass difference of $\Lambda_c$ and its excitation
$\Lambda'_c$, $U_{P-wave} \approx m_{\Lambda'_c} - m_{\Lambda_c}
=310$ MeV,  from $\Theta_{c}$, and adding a constituent strange
quark contribution $\Delta_s = m_{\Xi_c} - m_{\Lambda_c} \approx
184$ MeV, Ref.\cite{8} obtained 2580 MeV for $U_c,D_c$ masses
Assuming the same $U_{P-wave}$ and $\Delta_s$ for beauty heavy
pentaquarks, $U_b, D_b$ masses are estimated to be 5920
MeV\cite{8}.

The degeneracy of mass for the particles in a multiplet is lifted
by quark mass differences, $m_u$, $m_d$ and $m_s$. The mass terms,
up to linear corrections in light quark masses, are given by

\begin{eqnarray}
L &=& m_0^{R_{c,b}}Tr(\bar R_{c,b} R_{c,b}) + \alpha_m^{R_{c,b}}
Tr[\bar R_{c,b}( M
+ M^\dagger) R_{c,b}]\nonumber\\
&+& m_0^{S_{c,b}}Tr(\bar S_{c,b} S_{c,b}) + \alpha_m^{S_{c,b}}
Tr[\bar S_{c,b}( M
+ M^\dagger) S_{c,b}]\nonumber\\
\end{eqnarray}
We have neglected terms of the form $Tr[\bar R(\bar S) R(S)]Tr(M)$
which only re-scales $m_0$. $M$ is the quark mass matrix and is
given by

\begin{eqnarray}
M = \left ( \begin{array}{lll}
m_u&0&0\\
0&m_d&0\\
0&0&m_s
\end{array}
\right ).
\end{eqnarray}

Neglecting small $m_{u,d}$ masses, we obtain

\begin{eqnarray}
&&m_{U_{c,b}} = m_{D_{c,b}} = m_0^{R_{c,b}},\;\; m_{S_{c,b}} =
m_0^{R_{c,b}} + 2 \alpha^{R_{c,b}} m_s,\nonumber\\
&&m_{T_{c,b}} = m_0^{S_{c,b}},\;\;
m_{N_{c,b}} = m_0^{S_{c,b}} + \alpha^{S_{c,b}}_m m_s,\;\;
m_{\Theta_{c,b}} = m^{S_{c,b}}_0 + 2 \alpha^{S_{c,b}}_m m_s.
\end{eqnarray}

Taking into account of the $SU(3)_f$ breaking effects by
differences in light quark masses from
$\Delta_s=m_{\Xi_c}-m_{\Lambda_c}$ for constituent strange quark,
the masses of $S^{-}_{c}$ and $S^0_b$ were estimated to be 2770
MeV and 6100 MeV, respectively in Ref.\cite{8}. Making the same
assumption we obtain the masses of $N_c$, $N_b$, $T_c$ and $T_b$
to be 2894 MeV, 6236 MeV, 3078 MeV and 6420 MeV, respectively.
These values are similar to the estimates obtained in
Ref.\cite{8a}.

There are other model estimates for heavy pentaquark masses which
give larger masses. For example in the model of Karliner and
Lipkin, where the pentaquarks are formed from a triquark and a
diquark bound states\cite{4}, the masses are estimated to be 2985
MeV and 6398 MeV, respectively, which are above strong $pD$ and
$nB$ decay threshold. And the masses of $N_c$, $N_b$, $T_c$ and
$T_b$ are estimated to be 3165 MeV, 6570 MeV, 3340 MeV and 6740
MeV, respectively\cite{8a}. Removing the P-wave excitation
energy\cite{4} $\delta E^{P-wave} \approx 207$ between the diquark
and triquark system from $\Theta_{c,b}$ and adding the mass
difference due to the replacement of a light $u$ or $d$ quark by
an $s$ quark, one obtains the masses of $U^0_{c},\;D^-_c$ and
$U^+_{b},\; D^0_b$ to be 2858 MeV and 6533 MeV, respectively.
$S^-_c$ and $S^0_b$ are approximately 3028 MeV and 6708 MeV.
Clearly the above estimates for the masses are rather rough and
should not be expected to hold to more than to within 50 MeV or
even 100 MeV.

If the H1 narrow resonance of mass 3099 MeV is indeed the
$\Theta_c$ particle, both the diquark and triquark-diquark model
predictions for the mass are slightly lower than data. There is
also the possibility that the narrow resonant state observed at H1
is a chiral partner of the $\Theta_c$.  At present the
uncertainties involved in the estimates are large, it is too early
to make a decisive conclusion. With a mass 3099 MeV for
$\Theta_c$, it is possible for it to decay into $D^{*-} p$ and
$D^+ \bar p$. Similar situation may happen for beauty pentaquarks.
We therefore will consider processes involving both $D$, $B$ and
$D^*$, $B^*$.

We now write down the strong decay amplitudes to the leading order
using $SU(3)_f$ symmetry for heavy pentaquark decays with a $B$ or
a $D$ in the final states. We have

\begin{eqnarray}
&&L_{R_bNB} = c_{R_b NB} \bar R_b^jN_j^i \bar B_i +H.C.\nonumber\\
&&L_{S_bNB} = c_{S_b NB} \bar S_{b,jk} N^j_l \bar B_i \epsilon^{ikl}
+ H.C.\nonumber\\
&&L_{R_cND} = c_{R_c ND} \bar R_c^j N_j^i\bar D_i +H.C.\nonumber\\
&&L_{S_cND} = c_{S_c ND} \bar S_{c,jk} N^j_l \bar D_i
\epsilon^{ikl} + H.C.. \label{strong}
\end{eqnarray}

In the above $N$ is the ordinary baryon octet, $D^i$ and $B^i$ are
the charm and beauty mesons. They are given by

\begin{eqnarray}
&&N = (N_i^j) = \left ( \begin{array}{lll}
{\Sigma^0\over \sqrt{2}} + {\Lambda\over \sqrt{6}}&\Sigma^+&p\\
\Sigma^-&-{\Sigma^0 \over \sqrt{2}} + {\Lambda\over \sqrt{6}}&n\\
\Xi^-&\Xi^0&-{2\Lambda\over \sqrt{6}}
\end{array}
\right ).\nonumber\\
&&D = (D^i) = (D_u, D_d, D_s) = (c \bar u, c\bar d, c\bar
s),\nonumber\\
&&B = (B^i) = (B_u, B_d,B_s) = (b\bar u, b \bar d, b\bar s).
\end{eqnarray}

In Tables \ref{RNB1} and \ref{SNB}, we list the couplings for
$\bar N R_b (S_b) B$. One can obtain the couplings for $\bar N R_c
(S_c) D$ by replacing $B_i$ by $D_i$ and the sub-index $b$ to $c$.
Through out the paper, in equations for pentaquark couplings only
group indices are properly labelled and all fields in the
Lagrangian are going outwards. The Lorentz structures are
suppressed in the equations, and the proper ones are given in the
Tables. Also we will assume that heavy pentaquarks are spin-1/2
particles. To obtain results for spin-3/2 heavy pentaquarks, one
just uses at an appropriate place the Rarita-Schwinger
vector-spinor for the relevant fields.

We would like to point out that the above equations can be equally
applied to processes with $B$ and $D$ replaced by $D^*$ and $B^*$,
respectively. We will not distinguish them in equations in later
discussions unless specifically indicated.

\begin{table}[htb]
\caption{ Couplings for $B\bar N R_b (D\bar N R_c)$ in unit
$c_{RNB} (c_{RND})$. The Lorentz structure for the bi-spiner
product is of the form $\bar N \Gamma_P R$ with $\Gamma_P = +1$
and $\gamma_5$ for negative and positive parity for $R$,
respectively. For $B^* \bar N R_b$, the Lorentz structure for the
bi-spiner product should be changed to $\bar N \gamma_\mu
\gamma_5\Gamma_P R_b$.}\label{RNB1} \vspace{0.5cm}
\begin{tabular}{|l|l||l|l|}
 \hline
 $B_u$&$({1\over \sqrt{6}} \bar \Lambda + {1\over {2}}
\bar \Sigma^0)U_b^+ + \bar \Sigma^- D_b^0+ \bar \Xi^- S_b^0$&

$D_u$&$({1\over \sqrt{6}} \bar \Lambda + {1\over {2}} \bar
\Sigma^0)U_c^0 + \bar \Sigma^- D_c^-+ \bar \Xi^- S_c^-$
\\\hline

$B_d$&$\bar \Sigma^+  U_b^+ +  ({1\over \sqrt{6}} \bar \Lambda -
{1\over \sqrt{2}}\bar \Sigma^0) D^0_b +  \bar \Xi^0 S_b^0$&
$D_d$&$\bar \Sigma^+  U_c^0 +  ({1\over \sqrt{6}} \bar \Lambda -
{1\over \sqrt{2}}\bar \Sigma^0) D^-_c +  \bar \Xi^0 S_c^-$

\\\hline

$B_s$&$\bar p U^+_b+  \bar n D_b^0- \sqrt{{2\over 3}} \bar \Lambda
S^0_b$&

$D_s$&$\bar p U^0_c+  \bar n D_c^- - \sqrt{{2\over 3}} \bar
\Lambda S^-_c$

\\\hline
\end{tabular}
\end{table}

\begin{table}[htb]
\caption{ Couplings for $B \bar N S_b (D\bar N S_c)$ in unit
$c_{SNB} (c_{SND})$. Lorentz structure is the same as in Table
\ref{RNB1}.}\label{SNB} \vspace{0.5cm}
\begin{tabular}{|l|l|}
\hline
 $B_u$&${1\over 2} [\sqrt{2} \bar  \Xi^- T^0_b + 2 \bar
\Xi^0 T^+_b- \sqrt{2} \bar  \Sigma^- N_b^0-\sqrt{3} \bar \Lambda
N^+_b+ \bar  \Sigma^0 N^+_b- 2\bar  n \Theta^+_b ]$\\\hline
$B_d$&${1\over 2} [- 2\bar  \Xi^- T^-_b - \sqrt{2} \bar \Xi^0
T^0_b + \sqrt{2} \bar  \Sigma^+ N^+_b +\sqrt{3} \bar  \Lambda
N^0_b+ \bar  \Sigma^0 N^0_b+ 2\bar  p \Theta^+_b]$\\\hline
$B_s$&${1\over 2}[2\bar  \Sigma^- T^-_b- 2\bar  \Sigma^0 T^0_b - 2
\bar  \Sigma^+ T^+_b- \sqrt{2} \bar  p N^+_b+ \sqrt{2} \bar n
N^0_b]$
\\\hline

$D_u$&${1\over 2} [\sqrt{2} \bar  \Xi^- T^-_c + 2 \bar \Xi^0
T^0_c- \sqrt{2} \bar  \Sigma^- N_c^--\sqrt{3} \bar \Lambda N^0_c+
\bar \Sigma^0 N^0_c- 2\bar  n \Theta^0_c ]$\\\hline

$D_d$&${1\over 2} [- 2\bar  \Xi^- T^{--}_c - \sqrt{2} \bar \Xi^0
T^-_c + \sqrt{2} \bar  \Sigma^+ N^0_c +\sqrt{3} \bar  \Lambda
N^-_c+ \bar  \Sigma^0 N^-_c+ 2\bar  p \Theta^0_c]$\\\hline

$D_s$&${1\over 2}[2\bar  \Sigma^- T^{--}_c- 2\bar  \Sigma^0 T^-_c
- 2 \bar  \Sigma^+ T^0_c- \sqrt{2} \bar  p N^0_c+ \sqrt{2} \bar n
N^-_c]$
\\\hline

\end{tabular}
\end{table}

If the diquark model for pentaquarks is the right one, we see that
all the strong decay modes are forbidden due to restriction of
phase space. However, if the masses are close to the
triquark-diquark model predictions, strong decays are allowed. The
H1 data indicate that the above strong decays are possible. One
can use the allowed decay modes to determine the parameter
$c_{abc}$ and therefore the widths of the heavy pentaquarks. Here
we give the formula for the couplings in terms of decay widths
assuming the decays are allowed. For decays with $B$ in the final
states, we have,

\begin{eqnarray}
&&c^2_{R_bNB} = {16 \pi m_{S^0_b} \Gamma(S^0_b \to \Xi^- \bar B_u)
\over [(\hat \pi m_{S^0_b} + m_{\Xi^-})^2 - m_{B_u}^2]
Ph(m_{S^0_b},m_{\Xi^-},
m_{B_u})},\nonumber\\
 &&c^2_{S_bNB} = {16 \pi m_{\Theta^+_b}\Gamma(\Theta^+_b \to p \bar B_d)\over
[(\hat \pi m_{\Theta^+_b} m_{p})^2 - m_{B_d}^2]
Ph(m_{\Theta^+_b},m_p, m_{B_d})},
\end{eqnarray}
where $Ph(a,b,c) = \sqrt{1-(b+c)^2/a^2)(1-(b-c)^2/a^2)}$. $\hat
\pi$ is the eigenvalue of parity of the heavy pentaquark.

For decays with $B^*$ in the final states, we have

\begin{eqnarray}
&&c^2_{R_bNB^*} = {16 \pi m_{S^0_b} \Gamma(S^0_b \to \Xi^- \bar
B_u^*) \over f(m_{S^0_b},m_{\Xi^-}, m_{B_u^*})
Ph(m_{S^0_b},m_{\Xi^-},
m_{B_u^*})},\nonumber\\
&&c^2_{S_bNB^*} = {16 \pi m_{\Theta^+_b}\Gamma(\Theta^+_b \to p
\bar B_d^*)\over f(m_{\Theta^+_b},m_p, m_{B_d*})
Ph(m_{\Theta^+_b},m_p, m_{B_d*})},
\end{eqnarray}
where $f(a,b,c) =a^2+ b^2-c^2 -\hat \pi 6 a b +
((a^2-b^2)^2-c^4)/c^2$.

 Similarly one can obtain $c^2_{R_c(S_c)ND(D^*)}$ by
considering $S^-_c \to \bar D^0_u  (D^{*0}_u) \Xi^-$ and
$\Theta^0_c \to p \bar D_d^+ (\bar D_d^{*+})$ decays,
respectively.

At present there is only some information on the width of
$\Gamma(\Theta_c \to p \bar D^+_d)$. Assuming the narrow resonant
state of width $12\pm 3$ MeV at H1 is the $\Theta_c$ particle, we
obtain

\begin{eqnarray}
c^2_{S_c N D^*} \approx \left \{ \begin{array}{l}
1.712\;\;\;\;\hat \pi
= 1\\
0.167\;\;\;\;\hat \pi = -1.
\end{array}
\right .
\end{eqnarray}
Using these numbers, the decay widths for decay modes in Table
\ref{SNB} involving $D^*$ can be predicted. These predictions can
be tested.

\section{Weak Hadronic Decays of Heavy Pentaquarks}

Heavy pentaquark can also decay through weak interactions. If
kinematically the strong decays discussed in the previous section
are not allowed, weak interaction will dominate heavy pentaquark
decays. These decays can be semi-leptonic or purely hadronic ones.
Analysis on some of the heavy pentaquark properties have been
carried out\cite{6,6a,8,8a,9,10}. Here we will concentrate on some
two body hadronic heavy pentaquark decays.

\subsection {\bf $R_b(S_b) \to R_c(S_c) + \Pi$ decays}

In this subsection we study pentaquark decays of the type $R_b
(S_b) \to R_c(S_c) \Pi$. Here $\Pi$ represents a meson in the
pesudoscalar octet which is given by

\begin{eqnarray}
&&\Pi = (\Pi_i^j) = \left ( \begin{array}{lll}
{\pi^0\over \sqrt{2}} + {\eta\over \sqrt{6}}&\pi^+&K^+\\
\pi^-&-{\pi^0 \over \sqrt{2}} + {\eta\over \sqrt{6}}&K^0\\
K^-&\bar K^0&-{2\eta\over \sqrt{6}}
\end{array}
\right ).
\end{eqnarray}

The quark level effective Hamiltonian for $R_b (S_b) \to R_c (S_c)
\Pi$ is given by

\begin{eqnarray}
&&H_{eff} = {G_F\over \sqrt{2}} [V_{cb}^*V_{uq} (c_1 O_1 + c_2
O_2)
+ V_{ub}V_{cq}^* (c_1\tilde O_1 +c_2\tilde O_2)], \\
&&O_1 =\bar b \gamma_\mu (1-\gamma_5) c \bar u \gamma^\mu
(1-\gamma_5) q,\;\;O_2 =\bar b \gamma_\mu (1-\gamma_5) q \bar u
\gamma^\mu (1-\gamma_5) c,\\
&&\tilde O_1 =\bar b \gamma_\mu (1-\gamma_5) u \bar c \gamma^\mu
(1-\gamma_5) q,\;\;\tilde O_2 =\bar b \gamma_\mu (1-\gamma_5) q
\bar c \gamma^\mu (1-\gamma_5) u. \label{hamil}
\end{eqnarray}

The two operators $O_{1,2}$ in the above can induce decays of the
type, $R_b (S_b) \to R_c(S_c) + \Pi$, while the operators $\tilde
O_{1,2}$ will not cause beauty heavy pentaquark to charmed
pentaquark transitions. We write it down here for later
discussions.

Under $SU(3)_f$ symmetry $O_{1,2}$ transforms as an octet $H^i_j$.
With proper normalization the non-zero entries of $H^i_j$ can be
written as

\begin{eqnarray}
&& q = d,\;\;H^1_2 = 1;\;\;\;\; q = s\;\; H^1_3=1.
\end{eqnarray}

The $SU(3)_f$ invariant decay amplitudes are

\begin{eqnarray}
H(R_b \to R_c \Pi) &&= V^*_{cb}V_{uq}[r_{81} \bar R_{b,i}  R^i_c
\Pi^j_k H^k_j  + r_{82} \bar R_{bi}  R^j_c \Pi^k_j H^i_k +
r_{83}\bar R_{b,i} R_c^j \Pi^i_k H^k_j
];\nonumber\\
H(R_b\to S_c \Pi) &&=V^*_{cb}V_{uq}[s_{81} \bar R_{bi}  S_{c,jl}
\Pi^m_k H^l_m\epsilon^{ijk} +s_{82} \bar R_{bi} S_{c,jl} \Pi^i_m
H^l_k\epsilon^{jkm}
\nonumber\\
&&+s_{83} R_{b,i} \bar S_{c,jl} \Pi^l_m H^i_k\epsilon^{jkm}].
\label{eee}
\end{eqnarray}

The amplitudes for $S_b \to R_c \Pi$ can be obtained by
interchanging the indices $b$ and $c$, and treating the processes
as the charge conjugated ones in the second equation of
eq.(\ref{eee}) .

The amplitudes for $S_b \to S_c \Pi$ can be written as

\begin{eqnarray}
H(S_b \to S_c \Pi) &=& V_{cb}^*V_{uq}[s_1 \bar S_{b}^{ij} S_{c,ij}
\Pi^l_k H^k_l
+ s_2 \bar S_{b}^{ij} S_{c,kl} \Pi^i_kH^j_l\nonumber\\
&+& s_3 \bar S_{b}^{ij} S_{c,kj} \Pi^l_k H^i_l + s_4 \bar
S_{b}^{ij} S_{c,kj} \Pi^i_l H^l_k].
\end{eqnarray}

We list the results in Tables \ref{rr1}, \ref{rr2} and \ref{rr3}.
One can easily generalize the above formulation to the case with
the vector meson nonet $(\rho^{0,\pm}, K^{*,0,\pm}, \omega.
\phi)$.

\begin{table}[htb]
\caption{$SU(3)$ decay amplitudes for $ R_b \to R_c (S_c) \Pi$.
The Lorentz structure of the bi-spiner product is of the form
$\bar R_c (\bar S_c) (1+b\gamma_5) R_b$. Here $b$ is a
parameter.}\label{rr1} \vspace{0.5cm}
\begin{tabular}{|l|l||l|l|}
\hline
 $ U_b^+$ decay&$\Delta S =0$ &&$\Delta S = 1$\\\hline

$ U^0_c \pi^+$& $r_{81}+r_{83}$& $ U^0_c K^+$&
$r_{81}+r_{83}$\\\hline

$ T^0_c K^+$&$ s_{81}-s_{82}$&$ N^0_c K^+$& ${1\over \sqrt{2}}
(s_{81}-s_{82})$\\\hline

$ N^0_c \pi^+$&$- {1\over \sqrt{2}} (s_{81}-s_{82})$& $ \Theta^0_c
\pi^+$& $- (s_{81}-s_{82})$\\\hline

$ D^0_b$ decay&&&\\\hline

$ U^0_c  K^0$& ${1\over \sqrt{2}} (r_{82}-r_{83})$& $ U^0_c
\pi^0$& $r_{83}$\\\hline

$ U^0_c \eta$&$ {1\over \sqrt{6}} (r_{82}+r_{83})$ &$ D^-_c K^+
$&$ r_{81}$\\\hline

$ D^-_c \pi^+$& $r_{81}+ r_{82}$ &$ N^-_c K^+$& $-{1\over
\sqrt{2}}s_{81}$\\\hline

$ S^-_c K^+$&$ r_{82}$ &$ N^0_c  K^0$& $-{1\over
\sqrt{2}}s_{82}$\\\hline

$ T^-_c K^+$& $- {1\over \sqrt{2}}(s_{81}+s_{83})$ &$ \Theta^0_c
\pi^0$&$ {1\over \sqrt{2}}(s_{81}-s_{82})$\\\hline

$ T^0_c  K^0$&$ - (s_{82}+s_{83})$ &$ \Theta^0_c \eta$& ${1\over
\sqrt{6}}(s_{81}+s_{82})$\\\hline

$ N^-_c \pi^+$&$ {1\over \sqrt{2}}s_{83}$&&\\\hline

$ N^0_c \pi^0$& $ {1\over 2}(s_{81}-s_{82}-s_{83})$&&\\\hline

$ N^0_c \eta$& ${1\over 2\sqrt{3}}
(s_{81}+s_{82}+3s_{83})$&&\\\hline

$ \Theta^0_c \bar K^0$& $s_{83}$&&\\\hline

$ S^0_b$ decay&&&\\\hline

$ U^0_c \bar K^0$&$ r_{83}$& $ U^0_c \pi^0$& ${1\over \sqrt{2}}
r_{82}$\\\hline

$ S^-_c \pi^+$& $r_{81}$ &$ U^0_c \eta$& ${1\over \sqrt{6}}
(2r_{82}-r_{83})$\\\hline

$ T^-_c \pi^+$& ${1\over \sqrt{2}}s_{81}$ &$ D^-_c \pi^+$& $
r_{82}$\\\hline

$ T^0_c \pi^0$& $-{1\over \sqrt{2}}s_{81}$ & $ S^-_c K^+$
&$r_{81}+r_{82}$\\\hline

$ T^0_c \eta$& $-{1\over \sqrt{6}}(s_{81}-2s_{82})$ &$ T^-_c K^+$&
$- {1\over \sqrt{2}}s_{83}$\\\hline

$ N^0_c \bar K^0$& ${1\over \sqrt{2}}s_{82}$ &$ T^0_c  K^0$&$ -
s_{83}$\\\hline

&&$ N^-_c \pi^+$& ${1\over \sqrt{2}}(s_{81}+s_{83})$\\\hline

&&$ N^0_c \pi^0$& $-{1\over 2}(s_{81}+s_{83})$\\\hline

&&$ N^0_c \eta$&$ -{1\over 2\sqrt{3}}
(s_{81}-2s_{82}-3s_{83})$\\\hline

&&$ \Theta^0_c \bar K^0$& $s_{82}+s_{83} $
\\\hline
\end{tabular}
\end{table}

\begin{table}[htb]
\caption{$SU(3)$ decay amplitudes for $S_b \to S_c \Pi$ with
$\Delta S = 0$. The Lorentz structure is similar to Table
 \ref{rr1}.}\label{rr2} \vspace{0.5cm}
\begin{tabular}{|l|l|l|l|l|l|l|}
\hline
 $T^-_b$&$T^{--}_c \pi^+$&$T^-_c \pi^0$&$T^-_c \eta$&$T^0_c
\pi^-$ &$N^-_c \bar K^0$&$N^0_c K^-$\\\hline &$s_1 +s_3$&${1\over
2} (s_2 - s_3 + s_4)$&${1\over 2\sqrt{3}}
(s_2+s_3+s_4)$&$s_2$&${1\over \sqrt{2}}s_3$&${1\over
\sqrt{2}}s_2$\\\hline $T^0_b$&$T^-_c\pi^+$&$T^0_c\pi^0$&$T^0_c
\eta$&$N^0_c \bar K^0$&&\\\hline &${1\over 2}
(s_1+s_2+s_4)$&$-{1\over 2} s_2$&${1\over 2\sqrt{3}} s_2$
&${1\over 2}s_2$&&\\\hline $T^+_b$&$T^0_c \pi^+$&&&&&\\\hline
&$s_1+s_4$&&&&&\\\hline $N^0_b$&$T^-_c K^+$&$T^0_c K^0$ &$N^-_c
\pi^+$&$N^0_c \pi^0$&$N^0_c \eta$&$\Theta_c^0 \bar K^0$\\\hline
&${1\over 2}(s_2+s_4)$&${1\over \sqrt{2}}s_2$ &${1\over 2}
(2s_1+s_3)$ &$-{1\over 2\sqrt{2}}(s_3-s_4)$&$-{1\over
\sqrt{6}}(s_2 -s_3-s_4)$ &${1\over \sqrt{2}}s_3$\\\hline
$N^+_b$&$T^0_c K^+$&$N^0_c\pi^+$&&&&\\\hline &${1\over
\sqrt{2}}s_4$&${1\over 2}(2s_1+s_4)$&&&&\\\hline
$\Theta^+_b$&$N^0_c K^+$&$\Theta^0_c \pi^+$&&&&\\\hline &${1\over
\sqrt{2}} s_4$&$s_1$&&&&

\\\hline
\end{tabular}
\end{table}

\begin{table}[htb]
\caption{$SU(3)$ decay amplitudes for $S_b \to S_c \Pi$ with
$\Delta S = 1$. The Lorentz structure is similar to that in Table
\ref{rr1}.}\label{rr3} \vspace{0.5cm}
\begin{tabular}{|l|l|l|l|l|l|l|}
\hline
 $T^-_b$&$T^{--}_c K^+$&$T^-_c K^0$&$N^-_c \pi^0$&$N^-_c
\eta$ &$N^-_c \pi^-$&$\Theta^0_c K^-$\\\hline &$s_1 +s_3$&${1\over
\sqrt{2}}s_3$&${1\over 2} (s_2+s_4)$&${1\over
2\sqrt{3}}(s_2-s_3+s_4)$ &${1\over \sqrt{2}}s_2$&$s_2$\\\hline
$T^0_b$&$T^-_c K^+$&$N^-_c\pi^+$&$N^0_c
\pi^0$&$N^0_c\eta$&$\Theta^0_c \bar K^0$&\\\hline &${1\over
2}s_1$&${1\over 2} (s_2+s_4)$&$-{1\over 2\sqrt{2}} s_2$ &${1\over
2\sqrt{6}} s_2$&${1\over \sqrt{2}} s_2$&\\\hline $T^+_b$&$T^0_c
K^+$&$N^0_c \pi^+$&&&&\\\hline &$s_1$&${1\over
\sqrt{2}}s_4$&&&&\\\hline $N^0_b$&$N^-_c K^+$&$N^0_c K^0$
&$\Theta^0_c \pi^0$&$\Theta^0_c \eta$&&\\\hline &${1\over
2}(2s_1+s_2+s_3+s_4)$&${1\over 2}(s_2+s_3)$ &${1\over 2} s_4$
&$-{1\over \sqrt{3}}(s_2+s_3-s_4)$& &\\\hline $N^+_b$&$N^0_c
K^+$&$\Theta^0_c\pi^+$&&&&\\\hline &${1\over
2}(2s_1+s_4)$&${1\over \sqrt{2}}s_4$&&&&\\\hline
$\Theta^+_b$&$\Theta^0_c K^+$&&&&&\\\hline &$s_1+ s_4$&&&&&

\\\hline
\end{tabular}
\end{table}

\subsection {\bf $R_b (S_b) \to R_c (S_c) D (D^*)$}

The effective Hamiltonian for these processes is given by
\begin{eqnarray}
H_{eff} = {G_F\over \sqrt{2}}V^*_{cb}V_{cq}(c_1 O^c_1 + c_2
O^c_2).
\end{eqnarray}
Here $O^c_1= \bar b \gamma^\mu(1-\gamma_5) c \bar c \gamma_\mu
(1-\gamma_5) q$ and  $O^c_2= \bar c \gamma^\mu(1-\gamma_5) c \bar
b \gamma_\mu (1-\gamma_5)q$. In the above we have neglected small
penguin contributions. This effective Hamiltonian transforms as a
triplet $3$ with non-zero entries,

\begin{eqnarray}
&&q=d,\;\;H_2  = 1;\;\;\;\; q=s,\;\;H_3  = 1.
\end{eqnarray}

The $SU(3)_f$ invariant amplitudes can be written as
\begin{eqnarray}
H(R_b(S_b) \to R_c (S_c) D) &=& V_{cb}^*V_{cq}[r_{31} \bar R^i_b
R_{c,i} H_j D^j + r_{32}\bar R_b^i H_i R_{c,j} D^j + r
\epsilon_{ijk}\bar R^i_b S^{c,jl}D^kH_l\nonumber\\
 &+& s_{31}\bar
S_{b,ij}S^{ij}_c H_k D^k + s_{32} \bar S_{b,ij}S^{ik}H_kD^j].
\end{eqnarray}
The results for individual processes are given in Table \ref{dd}.

\begin{table}[htb]
\caption{$SU(3)$ decay amplitudes for $S_b \to S_c D$. The Lorentz
structure is similar to that in Table \ref{rr1}. For $S_b\to S_c
D^*$, the Lorentz structure of the bi-spiner product is of the
form $\bar S_c \gamma_\mu (1+b\gamma_5) S_b$.}\label{dd}
\vspace{0.5cm}
\begin{tabular}{|l|l||l|}
\hline $\Delta S = 0$&&$\Delta S = 1$\\\hline

$U^+_b$&$r_{31}U^0_c D_d$&$r_{31} U^0_c D_s$\\\hline

$D^0_b$&$(r_{31}+r_{32})D^-_c D_d + r_{32}U^0_c D_u + r_{32}S^-_c
D_s$&$r_{31}D^-_c D_s$\\\hline

$S^0_b$ &$r_{31} S^-_c D_d$&$(r_{31}+r_{32}) S^-_c D_s + r_{32}
U^0_c D_u + r_{31}D^-_c D_d$\\\hline

$U^+_b$&$rT^0_c D_s-{1\over \sqrt{2}} r N^0_c D_d$&${1\over
\sqrt{2}}r N^0_c D_s - r \Theta^0_c D_d$\\\hline

$D^0_b$&${1\over \sqrt{2}} r N^0_c D_u - {1\over \sqrt{2}}r T^-_c
D_s$&$-{1\over \sqrt{2}}r N^-_c D_s +r \Theta^0_c D_u$\\\hline

$S^0_b$&${1\over \sqrt{2}} r T^-_c D_d - r T^0_c D_u$&${1\over
\sqrt{2}}r N^-_c D_d - {1\over \sqrt{2}}r N^0_c D_u$\\\hline

$T^-_b$&$s_{31} T^{--}_c D_d + {1\over \sqrt{2}}s_{32} T^-_c
D_u$&$s_{31} T^{--}_c D_s + {1\over \sqrt{2}}s_{32} N^-_c
D_u$\\\hline

$T^0_b$&$(s_{31}+{1\over 2}s_{32})T^-_c D_d +{1\over
\sqrt{2}}s_{32} T^0_c D_u$&$s_{31} T^-_c D_s + {1\over 2}s_{32}
N^0_c D_u +{1\over 2}s_{32} N^-_c D_d$\\\hline

$T^+_b$&$(s_{31}+s_{32}) T^0_c D_d$&$s_{31} T^0_c D_s +{1\over
\sqrt{2}}s_{32}N^0_c D_d$\\\hline

$N^0_b$&$s_{31} N^-_c D_d + {1\over 2}s_{32}N^0_c D_u +{1\over
2}s_{32} T^-_c D_s$&$(s_{31}+{1\over 2}s_{32}) N^-_c D_s + {1\over
\sqrt{2}} s_{32}N^0_c D_u$\\\hline

$N^+_b$&$(s_{31}+{1\over 2}s_{32})N^0_c D_d + {1\over
\sqrt{2}}s_{32} T^0_c D_s$&$(s_{31}+{1\over 2}s_{32})N^0_c D_s +
{1\over \sqrt{2}}s_{32}\Theta^0_c D_d$\\\hline

$\Theta^+_b$&$s_{31}\Theta^0_c D_d + {1\over \sqrt{2}}s_{32} N^0_c
D_s$&$(s_{31}+s_{32})\Theta^0_c D_s$

\\\hline
\end{tabular}
\end{table}

\subsection {\bf $B \to N \bar R_c (\bar S_c)$ and $B\to \bar N
R_c(S_c)$}

The conjugate operators of $O_{1,2}$ in eq.(\ref{hamil}) is of the
form $\bar c b \bar q u$ and can induce decays of the type, $B \to
N + \bar R_c (\bar S_c)$.

The $SU(3)_f$ invariant decay amplitudes are

\begin{eqnarray}
H(R_c) &&= V_{cb}V_{uq}^*[\tilde r_{81} \bar B_i \bar R_c^j N^i_k
H^k_j + \tilde r_{82} \bar B_i \bar R^j_c N^k_j H^i_k
+ \tilde r_{83} \bar B_i \bar R^i_c N^j_k H^k_j];\nonumber\\
H(S_c) &&=V_{cb}V_{uq}^*[\tilde s_{81} \bar B_i \bar S_{cjl} N^m_k
H^l_m\epsilon^{ijk} +\tilde s_{82} \bar B_i \bar S_{cjl} N^i_m
H^l_k\epsilon^{jkm}
\nonumber\\
&&+\tilde s_{83} B_i \bar S_{cjl} N^l_m H^i_k\epsilon^{jkm}].
\end{eqnarray}

The conjugate operators in the second term of eq. (\ref{hamil}) is
of the form $\bar q b \bar u c$. This operator can induce $B \to
\bar N + R_c( S_c)$. It contains a $SU(3)$ triplet and an
anti-sixtet. The non-zero entries are:

\begin{eqnarray}
&&q =d,\;\;H(3c)_3 = 1,\;\;H(6c)^{12} = H(6c)^{21} = 1;\nonumber\\
&&q=s, \;\;H(3c)_2 = -1,\;\;H(6c)^{13} = H(6c)^{31} = 1.
\end{eqnarray}

One can write down $SU(3)_f$ decay amplitudes for $B\to \bar N R_c
(S_c)$ as the following

\begin{eqnarray}
H(R_c) &&= V_{ub}V_{cq}^*[\tilde r_{31} \bar B_i R_{cj} \bar N^i_l
H_k \epsilon^{jlk} + \tilde r_{32} \bar B_i R_{cj}\bar N^j_l
H_k\epsilon^{ilk}
\nonumber\\
&&+\tilde r_{61} \bar B_i R_{cj} \bar N^i_k H^{jk}
+\tilde r_{62} \bar B_i R_{cj} \bar N^j_k H^{ik}];\nonumber\\
H(S_c) &&= V_{ub}V_{cq}^*[\tilde s_{31} \bar B_i S_c^{ij} \bar
N^k_j H_k
+ \tilde s_{32} \bar B_i S_c^{jk}\bar N^i_j H_k\nonumber\\
&&+ \tilde s_{61} \bar B_i S_c^{ij}\bar N^k_l H^{lm}\epsilon_{jkm}
+\tilde s_{62} \bar B_i S_c^{jk} \bar N^l_k H^{im}\epsilon_{jlm}].
\end{eqnarray}

The hadronic parameters $\tilde r_{ij}$ are expected to be
similar, $\Gamma(B\to \bar N R_c(S_c))$ would be smaller by a
factor of $|V_{cb}V_{uq}^*|^2/|V_{ub}V_{cq}^*|^2$ compared with
$\Gamma(B \to N \bar R_c, (\bar S_c))$. The details are listed
 in Tables \ref{bc1},\ref{bc2} and \ref{bc3}.

\begin{table}[htb]
\caption{$SU(3)$ decay amplitudes for $B \to N \bar R_c (\bar
S_c)$. The Lorentz structure should be understood to be $\bar N(1
+ b\gamma_5)R_c(S_c)$ since weak interaction can have S- and
P-wave amplitudes.}\label{bc1} \vspace{0.5cm}
\begin{tabular}{|l|l||l|l|}
\hline
 $B_u$ decay &$\Delta S =0$ &&$\Delta S = -1$\\\hline

$\bar U^0_c \Sigma^-$& $\tilde r_{81}+\tilde r_{83}$& $\bar U^0_c
\Xi^-$& $\tilde r_{81}+\tilde r_{83}$\\\hline

$\bar T^0_c \Xi^-$&$ \tilde s_{81}-\tilde s_{82}$&$\bar N^0_c
\Xi^-$& ${1\over \sqrt{2}} (\tilde s_{81}-\tilde s_{82})$\\\hline

$\bar N^0_c \Sigma^-$&$- {1\over \sqrt{2}} (\tilde s_{81}-\tilde
s_{82})$& $\bar \Theta^0_c \Sigma^-$& $- (\tilde s_{81}-\tilde
s_{82})$\\\hline

$B_d$ decay&&&\\\hline

$\bar U^0_c \Sigma^0$& $-{1\over \sqrt{2}} (\tilde r_{81}-\tilde
r_{82})$& $\bar U^0_c \Xi^0$& $\tilde r_{81}$\\\hline

$\bar U^0_c \Lambda$&$ {1\over \sqrt{6}} (\tilde r_{81}+\tilde
r_{82})$ &$\bar D^-_c \Xi^- $&$ \tilde r_{83}$\\\hline

$\bar D^-_c \Sigma^-$& $\tilde r_{82}+ \tilde r_{83}$ &$\bar N^-_c
\Xi^-$& $-{1\over \sqrt{2}}\tilde s_{81}$\\\hline

$\bar S^-_c \Xi^-$&$ \tilde r_{82}$ &$\bar N^0_c \Xi^0$& $-{1\over
\sqrt{2}}\tilde s_{82}$\\\hline

$\bar T^-_c \Xi^-$& $- {1\over \sqrt{2}}(\tilde s_{81}+\tilde
s_{83})$ &$\bar \Theta^0_c \Sigma^0$&$ {1\over \sqrt{2}}(\tilde
s_{81}-\tilde s_{82})$\\\hline

$\bar T^0_c \Xi^0$&$ - (\tilde s_{82}+\tilde s_{83})$ &$\bar
\Theta^0_c \Lambda$& ${1\over \sqrt{6}}(\tilde s_{81}+\tilde
s_{82})$\\\hline

$\bar N^-_c \Sigma^-$&$ {1\over \sqrt{2}}\tilde s_{83}$&&\\\hline

$\bar N^0_c \Sigma^0$& $ {1\over 2}(\tilde s_{81}-\tilde
s_{82}-\tilde s_{83})$&&\\\hline

$\bar N^0_c \Lambda$& ${1\over 2\sqrt{3}} (\tilde s_{81}+\tilde
s_{82}+3\tilde s_{83})$&&\\\hline

$\bar \Theta^0_c n$& $\tilde s_{83}$&&\\\hline

$B_s$ decay&&&\\\hline $\bar U^0_c n$&$ \tilde r_{81}$& $\bar
U^0_c \Sigma^0$& ${1\over \sqrt{2}} \tilde r_{82}$\\\hline

$\bar S^-_c \Sigma^-$& $r_{83}$ &$\bar U^0_c \Lambda$& $-{1\over
\sqrt{6}} (2\tilde r_{81}-\tilde r_{82})$\\\hline

$\bar T^-_c \Sigma^-$& ${1\over \sqrt{2}}\tilde s_{81}$ &$\bar
D^-_c \Sigma^-$& $ \tilde r_{82}$\\\hline

$\bar T^0_c \Sigma^0$& $-{1\over \sqrt{2}}\tilde s_{81}$ & $\bar
S^-_c \Xi^-$ &$\tilde r_{82}+\tilde r_{83}$\\\hline

$\bar T^0_c \Lambda$& $-{1\over \sqrt{6}}(\tilde s_{81}-2\tilde
s_{82})$ &$\bar T^-_c \Xi^-$& $- {1\over \sqrt{2}}\tilde
s_{83}$\\\hline

$\bar N^0_c n$& ${1\over \sqrt{2}}\tilde s_{82}$ &$\bar T^0_c
\Xi^0$&$ - \tilde s_{83}$\\\hline

&&$\bar N^-_c \Sigma^-$& ${1\over \sqrt{2}}(\tilde s_{81}+\tilde
s_{83})$\\\hline

&&$\bar N^0_c \Sigma^0$& $-{1\over 2}(\tilde s_{81}+\tilde
s_{83})$\\\hline

&&$\bar N^0_c \Lambda$&$ -{1\over 2\sqrt{3}} (\tilde
s_{81}-2\tilde s_{82}-3\tilde s_{83})$\\\hline

&&$\bar \Theta^0_c n$& $\tilde s_{82}+\tilde s_{83} $
\\\hline
\end{tabular}
\end{table}

\begin{table}[htb]
\caption{$SU(3)$ decay amplitudes for $B \to \bar N (R_c, S_c)$
with $\Delta S = 0$. The Lorentz structure is of the form $\bar
R_c(\bar S_c) (1+b\gamma_5) N$.}\label{bc2} \vspace{0.2cm}
\begin{tabular}{|l|l|l|}
\hline

$B_u$&$\bar \Sigma^+ U^0_c$& $(\tilde r_{31}+\tilde r_{32} +
\tilde r_{61} + \tilde r_{62})$\\\hline

&$\bar \Sigma^0 D^-_c$& $-{1\over \sqrt{2}} (\tilde r_{31}+\tilde
r_{32} - \tilde r_{61} + \tilde r_{62})$\\\hline

&$\bar \Lambda D^-_c $& $-{1\over \sqrt{6}} (\tilde r_{31}-\tilde
r_{32}+2\tilde r_{33}-\tilde r_{61}-\tilde r_{62})$\\\hline

&$ \bar \Xi^0 S^-_c$& $\tilde r_{32}-\tilde r_{33}+\tilde
r_{62}$\\\hline

&$\bar \Xi^- T^{--}_c$& $\tilde s_{31}-\tilde s_{61}-\tilde
s_{62}$
\\\hline

&$\bar \Xi^0 T^-_c$& ${1\over \sqrt{2}} (\tilde s_{31}+\tilde
s_{61}-\tilde s_{62})$\\\hline

&$\bar \Sigma^0 N^-_c$& ${1\over 2}(\tilde s_{32}+2\tilde
s_{61}+\tilde s_{62})$\\\hline

&$\bar \Lambda N^-_c$& $-{1\over 2\sqrt{3}}(2\tilde s_{31}-\tilde
s_{32} -3\tilde s_{62})$\\\hline

&$\bar \Sigma^+ N^0_c$& ${1\over \sqrt{2}} (\tilde s_{32} + \tilde
s_{62})$\\\hline

&$\bar p \Theta^0_c$& $\tilde s_{32}+\tilde s_{62}$\\\hline

$B_d$& $\bar \Sigma^0 U^0_c$& $-{1\over \sqrt{2}}(\tilde
r_{31}+\tilde r_{32}+\tilde r_{61}-\tilde r_{62})$\\\hline

&$\bar \Lambda U^0_c$& ${1\over \sqrt{6}}(\tilde r_{31}-\tilde
r_{32}+2\tilde r_{33}+\tilde r_{61}+\tilde r_{62})$\\\hline

&$\bar \Sigma^- D^-_c$ &$-(\tilde r_{31}+\tilde r_{32} - \tilde
r_{61}-\tilde r_{62})$\\\hline

&$\bar \Xi^- S^-_c$& $-(\tilde r_{32}-\tilde r_{33}+\tilde
r_{62})$\\\hline

&$\bar \Xi^- T^{--}_c$& ${1\over \sqrt{2}}(\tilde s_{31} - \tilde
s_{61}+\tilde s_{62})$\\\hline

&$\bar \Xi^0 T^0_c$& $\tilde s_{31}+\tilde s_{61}+\tilde
s_{62}$\\\hline

&$\bar \Sigma^- N^-_c$& ${1\over \sqrt{2}} (\tilde s_{32}-\tilde
s_{62})$\\\hline

&$\bar \Lambda N^0_c$& $-{1\over 2\sqrt{3}} (2\tilde s_{31}-\tilde
s_{32}+3\tilde s_{62})$\\\hline

&$\bar \Sigma^0 N^0_c$& $-{1\over 2}(\tilde s_{32}-2\tilde
s_{61}-\tilde s_{62})$\\\hline

&$\bar n \Theta^0_c $& $\tilde s_{32}-\tilde s_{62}$\\\hline

$B_s$& $\bar \Xi^0 U^0_c$& $\tilde r_{31}+\tilde r_{33} + \tilde
r_{61}$\\\hline

&$\bar \Xi^- D^-_c$& $-(\tilde r_{31}+\tilde r_{33}- \tilde
r_{61})$
\\\hline

&$\bar \Xi^-N^-_c$& ${1\over \sqrt{2}}(\tilde s_{31}+\tilde s_{32}
- \tilde s_{61})$\\\hline

&$\bar \Xi^0 N^0_c$& $ {1\over \sqrt{2}}(\tilde s_{31}+\tilde
s_{32} + \tilde s_{61})$\\\hline

&$\bar \Sigma^0 \Theta^0_c$ &$\sqrt{2}\tilde s_{61}$\\\hline

&$\bar \Lambda\Theta^0_c$& $-\sqrt{{2\over 3}} (\tilde
s_{31}+\tilde s_{32})$
\\\hline
\end{tabular}
\end{table}

\begin{table}[htb]
\caption{$SU(3)$ decay amplitudes for $B \to \bar N (R_c, S_c)$
with $\Delta S = -1$. The Lorentz structure is the same as Table
\ref{bc2}.}\label{bc3} \vspace{0.2cm}
\begin{tabular}{|l|l|l|}
\hline

 $B_u$& $\bar p U^0_c$& $\tilde r_{31}+\tilde r_{32} + \tilde r_{61}
+\tilde r_{62}$\\\hline

&$\bar n D^-_c$& $\tilde r_{32}-\tilde r_{33}+\tilde
r_{62}$\\\hline

&$\bar \Sigma^0 S^-_c$& $-{1\over \sqrt{2}} (\tilde r_{31}+\tilde
r_{33} - \tilde r_{61})$\\\hline

&$\bar \Lambda S^-_c$& $-{1\over \sqrt{6}} (\tilde r_{31}+2\tilde
r_{32}-\tilde r_{33}-\tilde r_{61}+2\tilde r_{62})$\\\hline

&$\bar \Sigma^- T^{--}_c$& $-(\tilde s_{31}-\tilde s_{61}-\tilde
s_{62})$\\\hline

&$\bar \Sigma^0 T^-_c$& $-{1\over 2}(\tilde s_{31}-\tilde s_{32}
-\tilde s_{61}-2\tilde s_{62})$\\\hline

&$\bar \Lambda T^-_c$& $-{1\over 2\sqrt{3}}(\tilde s_{31}+\tilde
s_{32}+3\tilde s_{61})$\\\hline

&$\bar \Sigma^+ T^0_c$& $-(\tilde s_{32}+\tilde s_{62})$\\\hline

&$\bar n N^-_c$& $-{1\over \sqrt{2}}(\tilde s_{31}+\tilde
s_{61}-\tilde s_{62})$\\\hline

&$\bar p N^0_c$& $-{1\over \sqrt{2}}(\tilde s_{32}+\tilde
s_{62})$\\\hline

$B_d$& $\bar n U^0_c$& $(\tilde r_{31}+\tilde r_{33} + \tilde
r_{61})$\\\hline

&$\bar \Sigma^- S^-_c$ &$-(\tilde r_{31}+\tilde r_{33}- \tilde
r_{61})$
\\\hline

&$\bar \Sigma^- T^-_c$& $-{1\over \sqrt{2}}(\tilde s_{31}+\tilde
s_{32}-\tilde s_{61})$\\\hline

&$\bar \Sigma^0 T^0_c$ &${1\over \sqrt{2}}(\tilde s_{31}+\tilde
s_{32}-\tilde s_{61}) $\\\hline

&$\bar \Lambda T^0_c$& $-{1\over \sqrt{6}} (\tilde s_{31}+\tilde
s_{32}+3 \tilde s_{61})$\\\hline

&$\bar n N^0_c$& $-{1\over \sqrt{2}} (\tilde s_{31}+\tilde
s_{32}+\tilde s_{61})$\\\hline

$B_s$& $ \bar \Sigma^0 U^0_c$& $-{1\over \sqrt{2}}(\tilde
r_{32}-\tilde r_{33}-\tilde r_{62})$\\\hline

&$\bar \Lambda U^0_c$& $-{1\over \sqrt{6}}(2\tilde r_{31}+\tilde
r_{32}+r_{33}+2\tilde r_{61}-\tilde r_{62})$\\\hline

&$\bar \Sigma^- D^-_c$ &$-(\tilde r_{32}-\tilde r_{33} -\tilde
r_{62})$\\\hline

&$\bar \Xi^- S^-_c$& $-(\tilde r_{31}+\tilde r_{32}-\tilde
r_{61}-\tilde r_{62})$\\\hline

&$\bar \Xi^- T^-_c$& $-{1\over \sqrt{2}} (\tilde s_{32}-\tilde
s_{62})$\\\hline

&$\bar \Xi^0 T^0_c$& $-(\tilde s_{32}-\tilde s_{62})$\\\hline

&$\bar \Sigma^0 N^0_c$& ${1\over 2}(\tilde s_{31} - \tilde
s_{61}+\tilde s_{62})$\\\hline

&$\bar \Lambda N^0_c$& $-{1\over 2 \sqrt{3}} (\tilde
s_{31}-2\tilde s_{32}+3\tilde s_{61}+3\tilde s_{62})$
\\\hline

&$\bar \Sigma^- N^-_c$& $-{1\over \sqrt{2}} (\tilde s_{31} -
\tilde s_{61}+\tilde s_{62})$\\\hline

&$\bar n \Theta^0_c$& $-(\tilde s_{31}+\tilde s_{61}+\tilde
s_{62})$

\\\hline
\end{tabular}
\end{table}

\section{Discussions and Conclusions}

If the recently discovered state $\Theta^+$ is interpreted as
pentaquark bound state with an $\bar s$ and four light quarks,
heavy pentaquarks with the $\bar s$ replaced by a $\bar b$ or a
$\bar c$ should exist. H1 experiment at HERA-B has obtained some
evidences for $\Theta_c$. They form $SU(3)_f$ triplets $R_{c,b}$
and anti-sixtets $S_{c,b}$. These states can be further studied at
future collider experiments. $R_c$ and $S_c$ can also be produced
from $B$ decays at $B$-factories. If pentaquarks $R_{c,b}$ and
$S_{c,b}$ are kinematically allowed to decay through strong
interactions, one can use Tables I and II to relate different
decay widths, and test the model.

At present there is only some evidence for $\Theta_c \to p
D^{*-}p$. Using the decay width $\Gamma = 12$MeV obtained from H1,
we determine $c_{SND^*}^2$ to be $1.712$ and 0.167 for $\Theta_c$
with positive and negative parities, respectively. One expects
that $c^2_{SNB^*}$ to be similar to $c^2_{SND^*}$. Using Table
\ref{SNB}, we can obtain other decay widths which can be tested in
the future. We have nothing much to say about the size of the
couplings except that we expect the couplings $c_{R_bNB}$ and
$c_{S_bND}$ are about the same as $c_{R_cND}$ and $c_{S_cND}$,
respectively. If one extends $SU(3)_f$ to $SU(4)_f$, the strong
couplings can be related in principle to the ones involving just
light pentaquarks\cite{6a}. We will not consider this possibility
here.

The heavy pentaquarks can decay through weak interaction. We have
parameterized some of the two body hadronic decays in terms of
$SU(3)_f$ invariant amplitudes. From the Tables obtained we see
that there are several relations among different decay modes. For
example, for $\Delta S = 0$ processes of the type $S_b\to S_c
\Pi$, from Table \ref{rr3} we obtain

\begin{eqnarray}
\Gamma(T^-_b\to T^0_c \pi^-) &=&{ 2}\Gamma(T^-_b\to N^0_c K^-) ={
4} \Gamma(T^0_b\to T^0_c \pi^0)= {
12}\Gamma(T^0_b \to T^0_c \eta)\nonumber\\
 &=& {
4}\Gamma(T^0_b\to N^0_c \bar K^0) ={ 2}\Gamma(N^0_b \to
T^0_c K^0),\nonumber\\
\Gamma(N^+_b\to T^0_c K^+) &=& \Gamma(\Theta^+_b \to N^0_c K^+).
\end{eqnarray}
More relations can be read off from the Tables. These relations
can be used to study the properties of heavy pentaquarks and test
the model provided that the decays have substantial branching
ratios which requires knowledge about the size of the $SU(3)_f$
invariant amplitudes.

Theoretical calculations of the decay amplitudes are very
difficult since multi-quarks are involved. However for certain
decays, the structure is very simple and can be related to
experimentally measured modes, such as $\Theta^+_b \to \Theta^0_c
\pi^+$ can be related to $\Lambda_b \to \Lambda_c^+ \pi^-$.

In $\Theta^+_b \to \Theta^0_c \pi^+$ decay, the main contributions
is due to factorized matrix elements where the $\pi$ is emitted
from two light quarks in the effective Hamiltonian, and the
transition of $\Theta_b^+$ to $\Theta^0_c$ is due to the
transition of a $\bar b$ quark to a $\bar c$ quark in the
Hamiltonian, and the structure of the rest of the four light
quarks in the pentaquarks are basically preserved. Based on this
intuitive picture, Ref.\cite{10} relates $\Theta_b^+\to \Theta^0_c
\pi^+$ to $\Lambda_b \to \Lambda_c^+ \pi^-$ using heavy quark
effective theory, and concluded that the branching ratios for
these two processes are similar. The decay rate
$\Gamma(\Theta_b^+\to \Theta_c^0\pi^+)$ is estimated to be about
$2.5 \Gamma(B^0\to D^- \pi^-)$. This prediction can be tested at
future collider experiments.

From Table \ref{rr2}, we see that $\Theta_b^+\to \Theta_c^0 \pi^+$
is proportional to the $SU(3)_f$ invariant amplitude $s_1$, one
therefore can use the estimate in Ref.\cite{10} to obtain an
estimate for $s_1$. The invariant amplitudes $s_{2,3,4}$ involves
more complicated topology and is much harder to estimate. Although
it is difficult to know all decay amplitudes, knowing $s_1$ one
can make some useful predictions. The branching ratios for
processes in Tables \ref{rr2} and \ref{rr3} which just depend on
$s_1$ are therefore known. Up to mass splitting corrections in
phase space, we obtain

\begin{eqnarray}
&&\Gamma(T^0_b\to T^-_c K^+)\sim {1\over 4} {|V_{us}|^2\over
|V_{ud}|^2}\Gamma(\Theta_b^+\to
\Theta_c^0\pi^+),\nonumber\\
&&\Gamma(T^+_b\to T^0_c K^+)\sim {|V_{us}|^2\over
|V_{ud}|^2}\Gamma(\Theta_b^+\to \Theta_c^0\pi^+).
\end{eqnarray}

Using heavy quark effective theory, one can also relate several
other $SU(3)_f$ invariant amplitudes to $\Lambda_b\to \Lambda_c
\pi$ by realizing the fact that all amplitudes for the heavy
pentaquark transitions of the form $R_{b,i}R_c^i$,
$S_{b,ij}S^{ij}_c$ have similar factorization structure, and their
strength should also be similar. We therefore expect that $s_1
\sim r_{81} \sim r_{31}\sim s_{31}$.

One then has, up to mass splitting corrections in phase space, the
following relations

\begin{eqnarray}
\Gamma(\Theta_b^+\to \Theta_c^0\pi^+) &\sim& \Gamma(S^0_b\to S^-_c
\pi^+) \sim {|V_{ud}|^2\over |V_{us}|^2}\Gamma(D^0_b\to D^-_0
K^+)\nonumber\\
&\sim& {|V_{ud}|^2\over |V_{cd}|^2}(\Gamma(U^+_b\to U^0_c
D_d),\;\Gamma(S^0_b\to S^-_c
D_d))\nonumber\\
&\sim&{|V_{ud}|^2\over |V_{cs}|^2}(\Gamma(U^+_b\to U^0_c
D_s),\;\Gamma(U^+_b\to D^-_c D_s),\;\Gamma(S^0_b\to D^-_c D_d))\nonumber\\
&\sim& {|V_{ud}|^2\over |V_{cd}|^2}(\Gamma(T^-_b\to T^{--}_c
D_d),\;\Gamma(N^0_b\to N^-_c D_d),\;\Gamma(\Theta^+_b\to
\Theta^0_c
D_d))\nonumber\\
&\sim&{|V_{ud}|^2\over |V_{cs}|^2}(\Gamma(T^-_b\to T^{--}_c D_s)
,\;\Gamma(T^0_b\to T^-_c D_s),\;\Gamma(T^+_b\to T^0_c D_s)).
\end{eqnarray}
The above relations also hold for processes with $D$ replaced by
$D^*$.

Pentaquark properties can also be studied at $B$ factories. We
have studied $B\to N \bar R_c(\bar S_c)$ and $B\to \bar N
R_c(S_c)$ decays. From Tables \ref{bc1}, \ref{bc2} and \ref{bc3},
we see that there are several relations. For example

\begin{eqnarray}
&&\Gamma(B_u \to \bar U^0_c \Sigma^-) = {|V_{ud}|^2\over
|V_{us}|^2}
\Gamma(B_u \to \bar U_c \Xi^-),\nonumber\\
&&\Gamma(B_u \to \bar p U^0_c) = {|V_{ud}|^2\over |V_{us}|^2}
\Gamma(B_u \to \bar \Sigma^+ U^0_c).
\end{eqnarray}

 Should the heavy pentaquarks be discovered, these relations
can also provide important information. The decay amplitudes for
$B\to N \bar R_c (\bar S_c)$ and $B\to \bar N R_c (S_c)$ are
difficult to estimate. We are not able to provide any reliable
estimate, except that we expect them to be smaller than $B\to
N\bar \Lambda_c$ amplitudes.

Using the same formulation, one can also study $B$ decays into an
ordinary baryon $N$ and a light pentaquark, such as $\Theta^+$ in
the anti-decuplet. We however expect that the branching ratios to
be smaller than $B\to N\bar N$. Since $B\to N\bar N$ have small
branching ratios, it may be difficult to study $B\to N \bar
\Theta^+$ experimentally. This situation may change if one studies
light pentaquark decays of $B$ by three body decays, such as $B\to
D p(n) \Theta$. $B\to DK$ decay has a branching ratio of order a
few times $10^{-4}$. The K has a strong coupling to $p(n)\Theta^+$
which can be determined from $\Theta^+$ decay. With
$\Gamma_\Theta$ of $\Theta^+$ width to be about a MeV, one can
obtain a branching ratio as large as $10^{-6}$ which is within the
reach of near future B factories.
\\

\noindent{Acknowledgement}

We would like to thank H.-Y. Cheng and Shi-Lin Zhu for useful
discussions. We thank H.-Y. Cheng for informing us a paper by
Cheng, Chua and Hwang\cite{cheng} where some of the heavy
pentaquark decays discussed in this paper are also considered
before its publication. We also thank M.-L. Yan for bringing his
early work on pentaquark to our attentions. This work is partially
supported by grants from NSC and NNSF.

\end{document}